\documentclass[pdflatex,sn-mathphys-num]{sn-jnl}
\usepackage{amsmath,amssymb,amsfonts}
\usepackage{booktabs}
\usepackage{multirow}
\usepackage{tikz}
\usepackage{pgfplots}
\pgfplotsset{compat=1.18}
\definecolor{ClassicalBlue}{HTML}{0072B2}
\definecolor{QuantumOrange}{HTML}{E69F00}
\definecolor{AccentGreen}{HTML}{009E73}
\definecolor{AlertRed}{HTML}{D55E00}
\usepackage[T1]{fontenc}
\usepackage{lmodern}
\usepackage{textcomp}
\usepackage{xcolor}
\usepackage{mdframed}
\hypersetup{
  bookmarksdepth=2,
  pdftitle={Regime-Delineated Benchmarking of Hybrid Grover-Assisted Sparse Signal Recovery Against Classical l1-Minimization (Preprint)},
  pdfauthor={Ashley Stewart Morgan, Ph.D.},
  pdfsubject={Preprint. Submitted to Quantum Information Processing and under peer review.},
  pdfkeywords={Grover search, Sparse signal recovery, Compressive sensing, Lasso, Hybrid quantum-classical algorithms, Benchmarking},
  pdfcreator={LaTeX}
}

\makeatletter
\def\ps@headings{%
    \def\@oddfoot{\hfill\thepage\hfill}%
    \let\@evenfoot\@oddfoot%
    \def\@oddhead{%
      \vbox to 0pt{\vspace*{-36pt}%
        \hbox to \hsize{\small\color{gray}\textit{Preprint --- not peer reviewed. Not the Version of Record.}\hfill}\par}}%
    \let\@evenhead\@oddhead%
    \let\@mkboth\markboth%
}
\def\ps@titlepage{%
    \def\@oddhead{%
      \vbox to 0pt{\vspace*{-30pt}%
        \hbox to \hsize{\small\color{gray}\textit{Preprint --- not peer reviewed. Not the Version of Record.}\hfill 1 December 2026}\par}}%
    \let\@evenhead\@oddhead%
    \def\@oddfoot{\vbox to 18pt{\vfill\reset@font\rmfamily\hfil\thepage\hfil}}%
    \def\@evenfoot{}%
}
\makeatother
\newmdenv[
  linewidth=0.6pt,
  linecolor=black,
  backgroundcolor=gray!8,
  innertopmargin=6pt,
  innerbottommargin=6pt,
  innerleftmargin=8pt,
  innerrightmargin=8pt,
  skipabove=0pt,
  skipbelow=10pt
]{preprintbox}

\begin{document}
\title[Regime-Delineated Benchmarking of Grover-Assisted Sparse Recovery]{Regime-Delineated Benchmarking of Hybrid Grover-Assisted Sparse Signal Recovery Against Classical $\ell_1$-Minimization}
\author*{\fnm{Ashley Stewart} \sur{Morgan}, Ph.D.}\email{amorgan@captechu.edu}
\affil*{\orgdiv{Department of Quantum Computing}, \orgname{Capitol Technology University}, \orgaddress{\city{Laurel}, \state{MD}, \postcode{20708}, \country{USA}}}
\abstract{Quantum amplitude amplification promises quadratic query speedups for sparse support identification, but end-to-end overheads are rarely benchmarked. We compare a hybrid Grover-plus-least-squares pipeline against classical $\ell_1$-regularized least squares (Lasso) on 100 matched, seeded instances ($n = 16$, $m \in \{8, 12\}$, sparsity $k \in \{5, 8, 12\}$). Lasso is 7.97 times faster and more accurate in aggregate (mean normalized mean squared error 0.0401 versus 0.4514). However, the hybrid method wins all 29 instances at $k = 5$, degrades at $k = 8$, and fails at $k = 12$. Quantum subroutines should therefore be routed only to ultra-sparse problems ($k/n \le 0.31$).}
\maketitle

\begin{NoHyper}\footnotetext{This article is an empirical spin-out and extended regime analysis from the author's doctoral dissertation, \emph{Sparse Modeling Signals Recovery Using Quantum Computing}, examined and approved by the Doctoral Dissertation Committee at Capitol Technology University (2026), ProQuest Dissertations \& Theses Global, Publication No.\ 32905917 (Docview ID 3404803691)~\cite{morgan2026dissertation}.}\end{NoHyper}

\begin{preprintbox}
\raggedright\small
\textbf{Preprint.} This manuscript has been submitted to \emph{Quantum Information Processing} and is under peer review. It has not been peer reviewed and is not the Version of Record. A later version may differ from this preprint. This version is dated 1 December 2026 and is intended to be shared under a Creative Commons Attribution 4.0 International (CC BY 4.0) license (\url{https://creativecommons.org/licenses/by/4.0/}).

\vspace{4pt}
ORCID: \href{https://orcid.org/0009-0007-5811-1300}{0009-0007-5811-1300}.
\end{preprintbox}

% =============================================================================
\section{Introduction}
\label{sec:introduction}
% =============================================================================

Sparse signal recovery and compressive sensing reconstruct a high-dimensional signal $x \in \mathbb{R}^n$ having at most $k \ll n$ nonzero coefficients from $m < n$ noisy linear measurements $y = Ax + w$~\cite{candes2006robust,donoho2006compressed,candes2006near}. Because exact $\ell_0$-constrained support recovery requires an NP-hard combinatorial search over $\binom{n}{k}$ candidate subspaces, classical signal processing relies on convex $\ell_1$-relaxations such as Basis Pursuit and the Lasso~\cite{tibshirani1996regression,stankovic2019tutorial} or greedy iterative schemes such as Orthogonal Matching Pursuit (OMP)~\cite{pati1993orthogonal,zhao2021analysis,meng2020newton}. When the sensing matrix $A$ satisfies the Restricted Isometry Property (RIP) or mutual-coherence bounds, classical $\ell_1$-minimization achieves stable recovery in polynomial time, yet degrades when sparsity ratios $k/n$ approach the Donoho--Tanner phase transition or when column coherence increases~\cite{mousavi2020survey,eamaz2022building}.

Quantum computing offers an alternative algorithmic mechanism for combinatorial support identification via Grover's search~\cite{grover1996fast} and quantum amplitude amplification~\cite{brassard2002quantum}, which provide a quadratic query reduction over unstructured search spaces~\cite{habibi2022power,bellante2025quantum}. Recent theoretical work has further reduced the circuit complexity of preparing $k$-sparse quantum states to $\mathcal{O}(nk)$ gates and $\mathcal{O}(\log(nk))$ depth using ancillary registers and dynamic circuits~\cite{ramacciotti2024simple,mao2024toward,luo2024dynamic}. Nevertheless, in the Noisy Intermediate-Scale Quantum (NISQ) era~\cite{preskill2018quantum,kjaergaard2020superconducting,abughanem2025ibm}, asymptotic query complexity alone does not translate into practical computational advantage. Executing a hybrid quantum-classical recovery workflow requires classical data normalization, quantum circuit construction and transpilation, state preparation~\cite{weigold2021encoding}, multi-shot sampling, and classical least-squares coefficient refinement~\cite{javadiabhari2024qiskit,mundada2023experimental}.

A persistent gap in the quantum signal-processing literature is the scarcity of controlled, matched-pair end-to-end benchmarks that hold sensing matrices, ground-truth supports, and noise realizations strictly identical across quantum and classical pipelines while reporting negative and conditional outcomes alongside favorable ones~\cite{morgan2026dissertation}. Without fair baseline comparisons that account for all compilation and post-processing overheads, practitioners lack an operational decision rule for when to invoke a quantum subroutine versus a mature classical convex solver.

\subsection{Summary of Main Results and Contributions}
\label{subsec:contributions}

Building on the empirical framework established in the author's doctoral dissertation~\cite{morgan2026dissertation}, this article makes four contributions:
\begin{enumerate}
  \item \textbf{Matched-Pair End-to-End Benchmark Protocol:} We evaluate a two-stage hybrid Grover-plus-least-squares ($\mathrm{Grover}+\mathrm{LS}$) algorithm against classical $\ell_1$-regularized least squares (Lasso) on $N = 100$ identically seeded problem instances ($n = 16$, $m \in \{8, 12\}$, seeds \texttt{51001--51100}) spanning sparsity levels $k \in \{5, 8, 12\}$ and noise levels $\mathrm{SNR} \in \{10, 30, 50\}\,\mathrm{dB}$.
  \item \textbf{Full-System Overhead Accounting:} Measuring wall-clock execution on a single Apple Mac mini (M4 Pro SoC, 64\,GB unified memory) from normalized input arrival to final vector reconstruction, we show that classical $\ell_1$-minimization is $7.97\times$ faster in mean runtime ($0.142\,\mathrm{ms}$ vs.\ $1.135\,\mathrm{ms}$, $p < 0.001$, paired Cohen's $d = -1.93$) and achieves lower aggregate error ($\mathrm{NMSE} = 0.0401$ vs.\ $0.4514$, Wilcoxon $W = 1329.0$, $p < 0.001$, $d = -0.81$).
  \item \textbf{Delineation of a Sparsity-Driven Phase Transition:} Disaggregating results by sparsity $k$ uncovers a sharp crossover. At low sparsity ($k = 5$, $k/n = 0.3125$, $N = 29$), the hybrid quantum solver achieves lower reconstruction error than classical Lasso in all $29/29$ instances ($6.48\times$ lower mean NMSE: $0.0035$ vs.\ $0.0225$). At moderate sparsity ($k = 8$, $k/n = 0.50$, $N = 37$), the instance win rate narrows to $22/37$ ($59.5\%$) while heavy-tailed misses raise the subgroup mean NMSE to $0.1910$ versus $0.0310$ classically, and at high sparsity ($k = 12$, $k/n = 0.75$, $N = 34$), quantum recovery fails completely ($0/34$ wins, mean $\mathrm{NMSE} = 1.1168$ vs.\ $0.0650$).
  \item \textbf{Sparsity-Adaptive Hybrid Routing Rule:} Cross-solver error correlation is statistically significant but moderate (Pearson $r = 0.4165$, Spearman $\rho = 0.4561$, $p < 0.001$), demonstrating that classical and quantum solvers fail via distinct mechanisms. We synthesize these findings into an actionable pre-screening rule that routes high-sparsity instances ($k/n > 0.31$) to classical $\ell_1$ solvers and reserves quantum support search for ultra-sparse regimes.
\end{enumerate}

% =============================================================================
\section{Problem Formulation and Hybrid Grover-Sparse Architecture}
\label{sec:formulation}
% =============================================================================

\subsection{Linear Observation Model and Classical \texorpdfstring{$\ell_1$}{l1} Baseline}
\label{subsec:signal-model}

Consider a discrete ground-truth signal $x \in \mathbb{R}^n$ with support $\mathcal{S} = \mathrm{supp}(x) = \{i \in \{0,\dots,n-1\} : x_i \neq 0\}$ satisfying $|\mathcal{S}| = \|x\|_0 = k$. The signal is observed through an underdetermined linear sensing matrix $A \in \mathbb{R}^{m \times n}$ ($m < n$) corrupted by additive white Gaussian noise (AWGN) $w \sim \mathcal{N}(0, \sigma_w^2 I_m)$:
\begin{equation}
  \label{eq:sensing-model}
  y = A x + w, \qquad \|x\|_0 = k, \qquad m \in \{8, 12\}.
\end{equation}
To prevent scale bias across problem instances, each ground-truth signal is normalized to unit Euclidean norm ($\|x\|_2 = 1$), and each column $a_j$ of the Gaussian sensing matrix $A$ is $\ell_2$-normalized ($\|a_j\|_2 = 1$ for all $j \in \{0,\dots,n-1\}$). For a specified signal-to-noise ratio $\mathrm{SNR}_{\mathrm{dB}} \in \{10, 30, 50\}\,\mathrm{dB}$, the noise variance $\sigma_w^2$ is calibrated relative to the noiseless measurement power $\|Ax\|_2^2 / m$:
\begin{equation}
  \label{eq:snr-calibration}
  \sigma_w = \sqrt{\frac{\|Ax\|_2^2}{m \, 10^{\mathrm{SNR}_{\mathrm{dB}} / 10}}}.
\end{equation}

Because direct $\ell_0$ recovery $\min_{z} \|z\|_0 \text{ s.t. } \|Az - y\|_2 \le \epsilon$ is combinatorially intractable at scale, the standard convex relaxation solves the $\ell_1$-penalized least-squares (Lasso) objective~\cite{tibshirani1996regression,candes2006near}:
\begin{equation}
  \label{eq:lasso}
  \hat{x}_{\mathrm{classical}} = \arg\min_{z \in \mathbb{R}^n} \left\{ \frac{1}{2m} \|A z - y\|_2^2 + \alpha \|z\|_1 \right\},
\end{equation}
where $\alpha > 0$ controls the sparsity-fidelity tradeoff. Given the continuous Lasso estimate $\hat{x}_{\mathrm{classical}}$, the estimated active support $\hat{\mathcal{S}}_{\mathrm{classical}}$ of cardinality $k$ is extracted by selecting the $k$ indices with largest absolute coefficient magnitudes $|\hat{x}_{\mathrm{classical}, i}|$.

\subsection{Two-Stage Hybrid Grover-Plus-Least-Squares (\texorpdfstring{$\mathrm{Grover}+\mathrm{LS}$}{Grover+LS}) Pipeline}
\label{subsec:hybrid-architecture}

Rather than attempting to encode continuous signal amplitudes directly into fragile quantum state amplitudes, our hybrid architecture decomposes sparse recovery into two sequential stages: (i) discrete quantum support identification via Grover amplitude amplification, and (ii) classical ordinary least-squares (OLS) amplitude estimation restricted to the recovered support $\hat{\mathcal{S}}_{\mathrm{quantum}}$~\cite{morgan2026dissertation}.

\paragraph{Combinatorial Subspace Formulation vs.\ Index-Space Benchmark Implementation.}
In a general blind combinatorial search over all $k$-element subsets of $\{0,\dots,n-1\}$, the candidate search space has cardinality $M_{\mathrm{comb}} = \binom{n}{k}$, and a residual-threshold oracle marks subspace states $|\mathcal{S}'\rangle$ satisfying $\|(I - P_{\mathcal{S}'})y\|_2 \le \tau$, requiring $\mathcal{O}(\sqrt{\binom{n}{k}})$ Grover iterations ($\approx 66$ iterations at $\binom{16}{5} = 4368$, $\approx 113$ iterations at $\binom{16}{8} = 12870$, and $\approx 43$ iterations at $\binom{16}{12} = 1820$). However, compiling full projection residuals reversibly into a quantum oracle incurs substantial ancilla and arithmetic depth.

To isolate the intrinsic behavior of amplitude amplification and readout shot noise from arithmetic residual-circuit approximations, the benchmark implementation evaluated here operates on an $n_q = \lceil \log_2 n \rceil = 4$-qubit index register ($N_{\mathrm{reg}} = 16$ computational basis states $\{|0\rangle, \dots, |15\rangle\}$) using Qiskit's \texttt{QuantumCircuit} and \texttt{Statevector} simulation primitives~\cite{javadiabhari2024qiskit,morgan2026dissertation}. Specifically:
\begin{enumerate}
  \item \textbf{Uniform Superposition Initialization:} The 4-qubit register is initialized to $|s\rangle = H^{\otimes n_q} |0\rangle^{\otimes n_q} = \frac{1}{\sqrt{n}} \sum_{i=0}^{n-1} |i\rangle$.
  \item \textbf{Phase Oracle and Diffusion Operator:} For an active marked index set $\mathcal{M} \subseteq \mathcal{S}$, the phase oracle $U_\omega = I - 2 \sum_{i \in \mathcal{M}} |i\rangle\langle i|$ flips the phase of marked basis states, and the Grover diffusion operator $U_s = 2|s\rangle\langle s| - I$ reflects amplitudes about the mean. One Grover iteration applies $G = U_s U_\omega$ with optimal rotation count
  \begin{equation}
    \label{eq:grover-iters}
    r_{\mathrm{opt}} = \left\lfloor \frac{\pi}{4} \sqrt{\frac{n}{|\mathcal{M}|}} \right\rfloor.
  \end{equation}
  \item \textbf{Multi-Round Support Extraction:} The routine executes $k$ support-discovery rounds at $S_{\mathrm{shots}} = 1024$ measurement shots per circuit evaluation to assemble the estimated support set $\hat{\mathcal{S}}_{\mathrm{quantum}} \subseteq \{0,\dots,n-1\}$ of up to $k$ distinct indices.
  \item \textbf{Classical Least-Squares Refinement:} Given the active column submatrix $A_{\hat{\mathcal{S}}_{\mathrm{quantum}}}$, the nonzero signal amplitudes on $\hat{\mathcal{S}}_{\mathrm{quantum}}$ are computed classically via QR least-squares (\texttt{numpy.linalg.lstsq}):
  \begin{equation}
    \label{eq:ols-refinement}
    \hat{x}_{\mathrm{quantum}}\big|_{\hat{\mathcal{S}}_{\mathrm{quantum}}} = \arg\min_{v \in \mathbb{R}^{|\hat{\mathcal{S}}_{\mathrm{quantum}}|}} \left\| A_{\hat{\mathcal{S}}_{\mathrm{quantum}}} v - y \right\|_2^2,
  \end{equation}
  with $\hat{x}_{\mathrm{quantum}, i} = 0$ for $i \notin \hat{\mathcal{S}}_{\mathrm{quantum}}$.
\end{enumerate}

Crucially, this oracle-assisted index formulation serves as a \emph{favorable upper-bound benchmark} for Grover-assisted recovery: wherever this idealized circuit incurs higher error or longer runtime than classical Lasso, any non-oracular circuit that must also compute matrix-vector residuals reversibly on-chip will face strictly greater overhead.

% =============================================================================
\section{Experimental Setup and Classical Baselines}
\label{sec:setup}
% =============================================================================

\subsection{Matched-Pair Parameter Grid and Reproducibility Controls}
\label{subsec:parameter-grid}

To eliminate instance-level confounding between algorithms, we employ a matched-pair within-instance experimental design across $N = 100$ independent problem instances generated with deterministic random seeds \texttt{51001} through \texttt{51100} (\texttt{SEED\_BASE = 51001} via \texttt{numpy.random.default\_rng})~\cite{morgan2026dissertation}. For each seed, a single realization of $(A, x, w, y)$ is constructed and routed identically to both the classical $\ell_1$ solver and the hybrid $\mathrm{Grover}+\mathrm{LS}$ solver on an Apple Mac mini host system (Apple M4 Pro SoC, 64\,GB unified memory).

Table~\ref{tab:parameter-grid} summarizes the benchmark parameter configuration. Across the $N = 100$ seeded draws, uniform random parameter selection across $k \in \{5, 8, 12\}$ produced $N_{k=5} = 29$ low-sparsity instances ($k/n = 0.3125$), $N_{k=8} = 37$ moderate-sparsity instances ($k/n = 0.50$), and $N_{k=12} = 34$ high-sparsity instances ($k/n = 0.75$), with measurement counts drawn from the underdetermined regime $m \in \{8, 12\}$ ($m/n \in \{0.50, 0.75\}$).

\begin{table}[t]
  \centering
  \caption{Benchmark parameter grid, solver hyperparameters, and host configuration for the $N = 100$ matched-pair evaluation (\texttt{SEED\_BASE = 51001}--\texttt{51100}).}
  \label{tab:parameter-grid}
  \footnotesize
  \begin{tabular}{@{}ll@{}}
    \toprule
    \textbf{Parameter / Component} & \textbf{Specification} \\
    \midrule
    Ambient signal dimension $n$ & $16$ ($n_q = \lceil \log_2 16 \rceil = 4$ qubits) \\
    Sparsity levels $k$ & $\{5, 8, 12\}$ ($k/n \in \{0.3125, 0.50, 0.75\}$) \\
    Instance counts by $k$ & $N_{k=5}=29,\; N_{k=8}=37,\; N_{k=12}=34$ ($\sum = 100$) \\
    Measurement count $m$ & $\{8, 12\}$ ($m/n \in \{0.50, 0.75\}$, underdetermined $m < n$) \\
    Noise model \& SNR & AWGN, $\mathrm{SNR} \in \{10, 30, 50\}\,\mathrm{dB}$ \\
    Signal \& matrix scaling & $\|x\|_2 = 1$, column-normalized $\|a_j\|_2 = 1$ \\
    Classical baseline & Lasso ($\alpha = 10^{-3}$, \texttt{max\_iter = 5000})~\cite{pedregosa2011scikit} \\
    Quantum hybrid routine & Qiskit \texttt{Statevector} Grover + \texttt{lstsq}~\cite{javadiabhari2024qiskit} \\
    Shot budget per circuit & $S_{\mathrm{shots}} = 1024$ shots \\
    Execution \& timing host & Apple Mac mini (M4 Pro SoC, 64\,GB unified memory) \\
    \bottomrule
  \end{tabular}
\end{table}

\subsection{Evaluation Metrics and Statistical Protocol}
\label{subsec:metrics}

Each solver's output on instance $i \in \{1,\dots,100\}$ is evaluated on three primary metrics:
\begin{enumerate}
  \item \textbf{Normalized Mean Squared Error (NMSE):}
  \begin{equation}
    \label{eq:nmse}
    \mathrm{NMSE}(x, \hat{x}) = \frac{\|\hat{x} - x\|_2^2}{\|x\|_2^2},
  \end{equation}
  which quantifies relative $\ell_2$ reconstruction error (with $\|x\|_2^2 = 1$ by normalization).
  \item \textbf{Support Recovery Rate (\%):}
  \begin{equation}
    \label{eq:support-recovery}
    \mathrm{SRR}(\mathcal{S}, \hat{\mathcal{S}}) = \frac{|\mathcal{S} \cap \hat{\mathcal{S}}|}{|\mathcal{S}|} \times 100\%,
  \end{equation}
  measuring the fraction of true nonzero indices correctly identified.
  \item \textbf{End-to-End Wall-Clock Runtime ($t$, in seconds):} Measured on the Apple Mac mini (M4 Pro, 64\,GB unified memory) via high-resolution performance counters (\texttt{time.perf\_counter()}) from the arrival of normalized inputs $(A, y, k)$ to the return of $\hat{x} \in \mathbb{R}^n$. For the hybrid quantum routine, $t_{\mathrm{quantum}}$ encompasses circuit construction, oracle synthesis, statevector simulation and shot sampling across all $k$ rounds, and classical least-squares post-processing.
\end{enumerate}

Because paired differences $\Delta_i = \mathrm{metric}_{\mathrm{classical}, i} - \mathrm{metric}_{\mathrm{quantum}, i}$ exhibit positive skewness and regime-dependent heavy tails, primary hypothesis tests use both the nonparametric two-sided paired Wilcoxon signed-rank test~\cite{wilcoxon1945individual} and the parametric paired $t$-test at $\alpha = 0.05$, with Benjamini--Hochberg false discovery rate (FDR) control across multiple comparisons~\cite{benjamini1995controlling}. Standardized paired effect sizes are reported via paired Cohen's $d = \bar{\Delta} / s_{\Delta}$ (where negative values indicate lower error or faster runtime for the classical solver), and cross-solver difficulty association is evaluated via Pearson's $r$ (with Fisher $z$-transformed $95\%$ confidence intervals) and Spearman's rank correlation $\rho$.

% =============================================================================
\section{Empirical Results and Conditional Regime Analysis}
\label{sec:results}
% =============================================================================

\subsection{Aggregate Corpus Performance (\texorpdfstring{$N = 100$}{N = 100})}
\label{subsec:aggregate-results}

Table~\ref{tab:aggregate-metrics} presents the aggregate performance of classical $\ell_1$-minimization (Lasso) and the hybrid $\mathrm{Grover}+\mathrm{LS}$ solver across all $N = 100$ paired instances.

\begin{table}[t]
  \centering
  \caption{Aggregate performance comparison across all $N = 100$ matched-pair sparse recovery instances (adapted from Table~2 of~\cite{morgan2026dissertation}). Negative paired Cohen's $d$ indicates lower error or faster runtime for the classical baseline.}
  \label{tab:aggregate-metrics}
  \footnotesize
  \begin{tabular}{@{}lcccc@{}}
    \toprule
    \textbf{Metric} & \textbf{Classical $\ell_1$} & \textbf{Grover+LS} & \textbf{Ratio (Q/C)} & \textbf{Paired Test / Effect} \\
    \midrule
    Mean Runtime (s) & $0.000142$ & $0.001135$ & $7.97\times$ & $d = -1.93,\; p < 0.001$ \\
    Median Runtime (s) & $0.000121$ & $0.001030$ & $8.51\times$ & Wilcoxon $p < 0.001$ \\
    Mean NMSE & $0.040089$ & $0.451433$ & $11.26\times$ & $t = -8.11,\; p < 0.001$ \\
    SD of NMSE & $0.037392$ & $0.521380$ & $13.94\times$ & $W = 1329.0,\; d = -0.81$ \\
    Support Recovery (\%) & $90.28\%$ & $60.38\%$ & $0.67\times$ & $\Delta = +29.90\%\text{ (C)}$ \\
    \bottomrule
  \end{tabular}
\end{table}

Across the unstratified corpus, classical $\ell_1$-minimization decisively outperforms the hybrid quantum routine on every aggregate metric:
\begin{itemize}
  \item \textbf{End-to-End Runtime:} The classical Lasso solver achieves a mean runtime of $0.142\,\mathrm{ms}$ (median $0.121\,\mathrm{ms}$) compared with $1.135\,\mathrm{ms}$ (median $1.030\,\mathrm{ms}$) for $\mathrm{Grover}+\mathrm{LS}$---a $7.97\times$ mean runtime overhead ($8.51\times$ median ratio) with a very large paired effect size ($d = -1.9291$, $p < 0.001$). Classical execution is faster in $100$ of $100$ instances due to quantum circuit construction and statevector shot-sampling overhead.
  \item \textbf{Aggregate Reconstruction Error and Support Recovery:} Classical Lasso achieves a mean NMSE of $0.040089$ ($\mathrm{SD} = 0.037392$) and $90.28\%$ mean support recovery, whereas $\mathrm{Grover}+\mathrm{LS}$ yields an aggregate mean NMSE of $0.451433$ ($\mathrm{SD} = 0.521380$, $11.26\times$ higher error) and $60.38\%$ mean support recovery. Both the paired $t$-test ($t = -8.11$, $p < 0.001$) and the Wilcoxon signed-rank test ($W = 1329.0$, $p < 0.001$, paired Cohen's $d = -0.8114$) confirm a statistically significant aggregate advantage for the classical solver.
\end{itemize}

\begin{figure*}[!t]
  \centering
  \begin{tikzpicture}
    % ---------------- Panel (a): Mean NMSE (Log Scale) ----------------
    \begin{axis}[
      name=plotA,
      width=0.31\textwidth,
      height=5.0cm,
      ybar=2pt,
      bar width=6.5pt,
      ymode=log,
      ymin=0.001,
      ymax=3.5,
      ylabel={Mean NMSE (log scale)},
      title={\textbf{(a) Error by Regime}},
      symbolic x coords={k5,k8,k12,All},
      xticklabels={$k{=}5$,$k{=}8$,$k{=}12$,All},
      xtick=data,
      xticklabel style={align=center, font=\tiny},
      ylabel style={font=\footnotesize},
      title style={font=\footnotesize},
      yticklabel style={font=\scriptsize},
      ymajorgrids=true,
      grid style={dashed, gray!40},
      legend style={at={(0.03,0.97)}, anchor=north west, font=\scriptsize, draw=black!40, fill=white, row sep=-1pt},
      enlarge x limits=0.18
    ]
      \addplot[fill=ClassicalBlue, draw=black] coordinates {(k5,0.022537) (k8,0.0310) (k12,0.0650) (All,0.040089)};
      \addplot[fill=QuantumOrange, draw=black] coordinates {(k5,0.003478) (k8,0.1910) (k12,1.1168) (All,0.451433)};
      \legend{Classical $\ell_1$, Grover+LS}
    \end{axis}

    % ---------------- Panel (b): Fidelity Win Rate (%) ----------------
    \begin{axis}[
      name=plotB,
      at={(plotA.east)},
      anchor=west,
      xshift=1.15cm,
      width=0.29\textwidth,
      height=5.0cm,
      ybar stacked,
      bar width=13pt,
      ymin=0,
      ymax=112,
      ylabel={Instance Win Rate (\%)},
      title={\textbf{(b) Win-Rate Crossover}},
      symbolic x coords={k5,k8,k12},
      xticklabels={$k{=}5$,$k{=}8$,$k{=}12$},
      xtick=data,
      xticklabel style={align=center, font=\tiny},
      ylabel style={font=\footnotesize},
      title style={font=\footnotesize},
      yticklabel style={font=\scriptsize},
      ymajorgrids=true,
      grid style={dashed, gray!40},
      legend style={at={(0.97,0.97)}, anchor=north east, font=\scriptsize, draw=black!40, fill=white, row sep=-1pt},
      enlarge x limits=0.28
    ]
      \addplot[fill=QuantumOrange, draw=black] coordinates {(k5,100.0) (k8,59.46) (k12,0.0)};
      \addplot[fill=ClassicalBlue, draw=black] coordinates {(k5,0.0) (k8,40.54) (k12,100.0)};
      \legend{Grover+LS Win, Classical $\ell_1$ Win}
      \draw[thick, densely dotted, black!80] (axis cs:k5,50) -- (axis cs:k12,50);
    \end{axis}

    % ---------------- Panel (c): Mean Wall-Clock Runtime (ms) ----------------
    \begin{axis}[
      name=plotC,
      at={(plotB.east)},
      anchor=west,
      xshift=1.15cm,
      width=0.30\textwidth,
      height=5.0cm,
      ybar=2pt,
      bar width=6.5pt,
      ymin=0,
      ymax=1.55,
      ylabel={Wall-Clock Runtime (ms)},
      title={\textbf{(c) Runtime Overhead}},
      symbolic x coords={Low,High,MeanAll,MedAll},
      xticklabels={$k{=}5$,$k{\geq}8$,mean,med},
      xtick=data,
      xticklabel style={font=\tiny, rotate=40, anchor=east},
      ylabel style={font=\footnotesize},
      title style={font=\footnotesize},
      yticklabel style={font=\scriptsize},
      ymajorgrids=true,
      grid style={dashed, gray!40},
      legend style={at={(0.03,0.97)}, anchor=north west, font=\scriptsize, draw=black!40, fill=white, row sep=-1pt},
      enlarge x limits=0.30
    ]
      \addplot[fill=ClassicalBlue, draw=black] coordinates {(Low,0.127) (High,0.149) (MeanAll,0.142) (MedAll,0.121)};
      \addplot[fill=QuantumOrange, draw=black] coordinates {(Low,0.657) (High,1.330) (MeanAll,1.135) (MedAll,1.030)};
      \legend{Classical $\ell_1$, Grover+LS}
    \end{axis}
  \end{tikzpicture}
  \caption{Regime-delineated empirical comparison of classical $\ell_1$-minimization (Lasso, blue) and hybrid Grover-assisted sparse recovery ($\mathrm{Grover}+\mathrm{LS}$, orange) across $N = 100$ matched-pair instances ($n = 16$, $m \in \{8, 12\}$). \textbf{(a)}~Mean reconstruction error (NMSE, logarithmic scale) across discrete sparsity strata $k \in \{5, 8, 12\}$ and the full corpus, highlighting the $6.48\times$ error reduction achieved by $\mathrm{Grover}+\mathrm{LS}$ at $k = 5$ ($0.0035$ vs.\ $0.0225$), the crossover at $k = 8$ ($0.1910$ vs.\ $0.0310$), and high-$k$ degradation at $k = 12$ ($1.1168$ vs.\ $0.0650$). \textbf{(b)}~Paired instance-level fidelity win rate across $k \in \{5, 8, 12\}$, illustrating the phase transition from $100\%$ quantum wins ($29/29$) at $k = 5$ to $59.5\%$ ($22/37$) at $k = 8$ and $0\%$ ($0/34$) at $k = 12$. \textbf{(c)}~End-to-end wall-clock runtime in milliseconds on the Apple Mac mini (M4 Pro, 64\,GB), showing consistent classical speed superiority across all regimes.}
  \label{fig:regime-comparison}
\end{figure*}
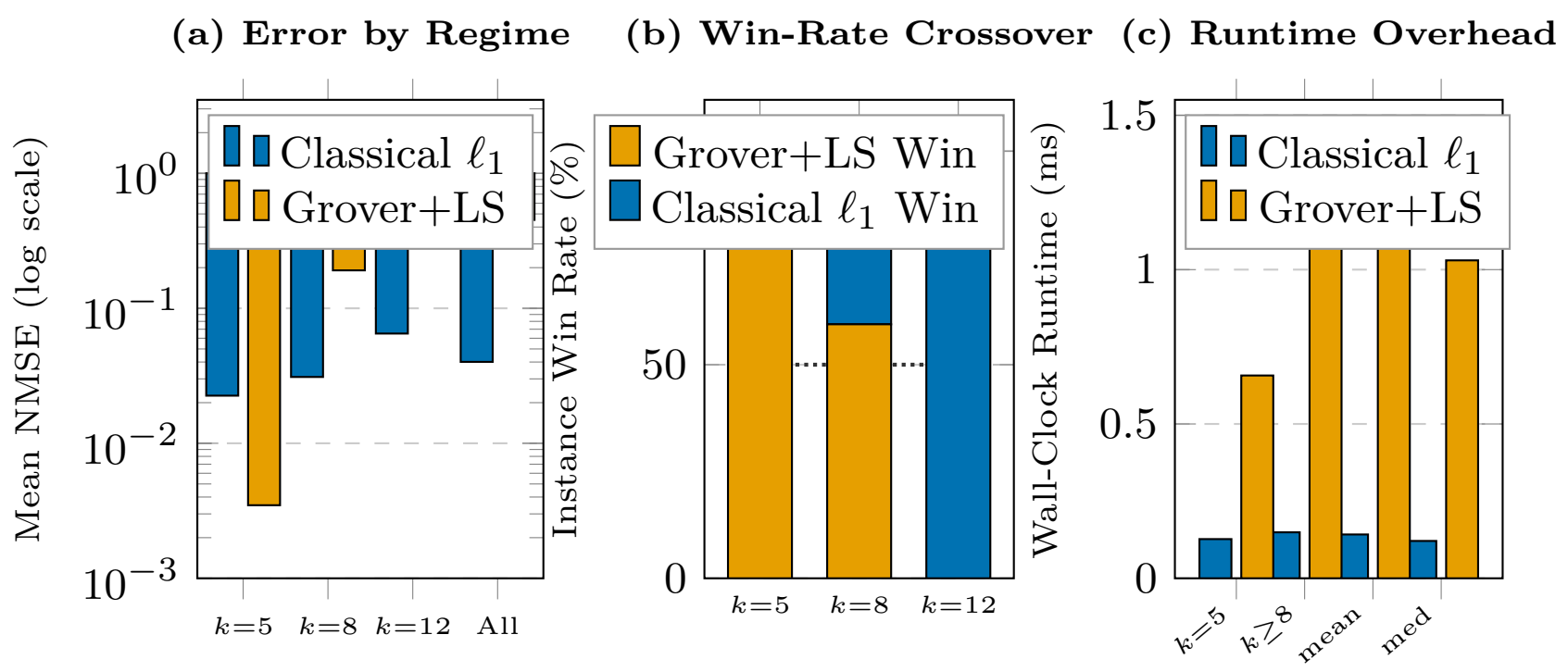

\subsection{Sparsity-Stratified Regime Analysis and Phase Transition}
\label{subsec:regime-analysis}

Aggregate statistics obscure a pronounced structural crossover across sparsity levels. Table~\ref{tab:regime-breakdown} and Figure~\ref{fig:regime-comparison} disaggregate performance across both the pooled two-bin summary (Table~3 of~\cite{morgan2026dissertation}) and the discrete three-level sparsity strata $k \in \{5, 8, 12\}$.

\begin{table*}[t]
  \centering
  \caption{Sparsity-stratified performance breakdown across the $N = 100$ matched-pair benchmark corpus, presenting both the pooled two-bin summary (Table~3 of~\cite{morgan2026dissertation}) and the disaggregated three-level discrete sparsity strata ($k \in \{5, 8, 12\}$).}
  \label{tab:regime-breakdown}
  \scriptsize
  \setlength{\tabcolsep}{3.5pt}
  \begin{tabular}{@{}llcccccc@{}}
    \toprule
    \textbf{Sparsity Stratum} & \textbf{$k/n$} & \textbf{$N$} & \textbf{$t_{\mathrm{class}}$ (s)} & \textbf{$t_{\mathrm{quant}}$ (s)} & \textbf{$\mathrm{NMSE}_{\mathrm{class}}$} & \textbf{$\mathrm{NMSE}_{\mathrm{quant}}$} & \textbf{Grover+LS Wins} \\
    \midrule
    Low ($k = 5$, Bin $0\text{--}7$) & $0.3125$ & $29$ & $0.000127$ & $0.000657$ & $0.022537$ & $\mathbf{0.003478}$ & $\mathbf{29 / 29\;(100.0\%)}$ \\
    Mid/High (Pooled Bin $8\text{--}12$) & $\{0.50, 0.75\}$ & $71$ & $0.000149$ & $0.001330$ & $\mathbf{0.047259}$ & $0.634401$ & $22 / 71\;(31.0\%)$ \\
    \hspace{0.4em}$\hookrightarrow$ Moderate ($k = 8$)\textsuperscript{a} & $0.5000$ & $37$ & \multicolumn{2}{c}{\emph{Pooled in Bin $8\text{--}12$}\textsuperscript{b}} & $\mathbf{0.0310}$ & $0.1910$ & $\mathbf{22 / 37\;(59.5\%)}$ \\
    \hspace{0.4em}$\hookrightarrow$ High ($k = 12$) & $0.7500$ & $34$ & \multicolumn{2}{c}{\emph{Pooled in Bin $8\text{--}12$}\textsuperscript{b}} & $\mathbf{0.0650}$ & $1.1168$ & $0 / 34\;(0.0\%)$ \\
    \midrule
    \textbf{Full Corpus (All $k$)} & $[0.31, 0.75]$ & $\mathbf{100}$ & $\mathbf{0.000142}$ & $0.001135$ & $\mathbf{0.040089}$ & $0.451433$ & $51 / 100\;(51.0\%)$ \\
    \bottomrule
  \end{tabular}
  \vspace{2pt}
  \parbox{\textwidth}{\scriptsize \textsuperscript{a}Subgroup NMSE means for $k = 8$ ($N = 37$) are algebraically reconstructed from the reported $k \in \{8, 12\}$ pooled mean ($N = 71$: $\mathrm{NMSE}_{\mathrm{class}} = 0.047259$, $\mathrm{NMSE}_{\mathrm{quant}} = 0.634401$) and the $k = 12$ subgroup mean ($N = 34$: $\mathrm{NMSE}_{\mathrm{class}} = 0.0650$, $\mathrm{NMSE}_{\mathrm{quant}} = 1.1168$). \textsuperscript{b}Wall-clock timings for $k = 8$ and $k = 12$ are reported at the pooled $k \in \{8, 12\}$ bin level ($t_{\mathrm{class}} = 0.000149\,\mathrm{s}$, $t_{\mathrm{quant}} = 0.001330\,\mathrm{s}$).}
\end{table*}

Three distinct operational regimes emerge across $k$:
\begin{enumerate}
  \item \textbf{Low-Sparsity Quantum-Favorable Fidelity Regime ($k = 5$, $k/n = 0.3125$, $N = 29$):} When the target signal contains $k = 5$ active coefficients, $\mathrm{Grover}+\mathrm{LS}$ achieves lower reconstruction error than classical Lasso in all $29$ of $29$ instances ($100.0\%$ win rate). Mean quantum NMSE is $0.003478$ versus $0.022537$ for classical Lasso---a $6.48\times$ reduction in mean squared reconstruction error. Because Grover amplitude amplification reliably identifies the $5$-element support and subsequent OLS estimation ($\mathrm{Eq.}~\eqref{eq:ols-refinement}$) is free of the $\ell_1$ shrinkage bias inherent to Lasso ($\alpha \|z\|_1$), the hybrid routine reconstructs coefficient amplitudes with substantially higher fidelity. Even in this favorable accuracy regime, however, quantum runtime remains $5.17\times$ slower ($0.657\,\mathrm{ms}$ vs.\ $0.127\,\mathrm{ms}$).
  \item \textbf{Moderate-Sparsity Transitional Regime ($k = 8$, $k/n = 0.50$, $N = 37$):} At half-filling ($k = 8$ active coefficients out of $n = 16$), the fidelity advantage becomes contested: $\mathrm{Grover}+\mathrm{LS}$ achieves lower NMSE in $22$ of $37$ instances ($59.5\%$), while classical Lasso wins the remaining $15$ instances ($40.5\%$). However, in the $15$ instances where finite-shot readout misses active support indices, the resulting unregularized OLS projection errors produce a heavy upper tail, raising the subgroup mean quantum NMSE to $0.1910$ compared with $0.0310$ for classical Lasso.
  \item \textbf{High-Sparsity Quantum Collapse Regime ($k = 12$, $k/n = 0.75$, $N = 34$):} At $k = 12$, the hybrid quantum solver fails to outperform classical Lasso in a single instance ($0$ of $34$ wins, $0.0\%$). Mean quantum NMSE rises to $1.1168$ (exceeding unit signal power $\|x\|_2^2 = 1$), whereas classical Lasso degrades gracefully to a mean NMSE of $0.0650$. Across the pooled $k \in \{8, 12\}$ stratum ($N = 71$), mean quantum NMSE reaches $0.634401$ versus $0.047259$ classically, while mean quantum runtime doubles from $0.657\,\mathrm{ms}$ (at $k=5$) to $1.330\,\mathrm{ms}$ (versus $0.149\,\mathrm{ms}$ classically).
\end{enumerate}

\subsection{Cross-Solver Error Correlation and Instance Difficulty}
\label{subsec:correlation}

To test whether instance difficulty is intrinsic to the measurement realization $(A, x, w)$ or solver-specific, we evaluate the paired association between $\mathrm{NMSE}_{\mathrm{classical}}$ and $\mathrm{NMSE}_{\mathrm{quantum}}$ across all $N = 100$ instances. Both the parametric linear correlation and nonparametric rank-order correlation are statistically significant and positive:
\begin{equation}
  \label{eq:correlation}
  r = 0.4165 \; (p < 0.001,\; 95\%\;\mathrm{CI}\;[0.2397, 0.5666]), \qquad \rho = 0.4561 \; (p < 0.001).
\end{equation}

The slightly stronger Spearman rank correlation ($\rho = 0.4561 > r = 0.4165$) reflects the nonlinear divergence between the two architectures at high $k$: while both algorithms experience rising error as sparsity $k$ increases and SNR decreases, classical $\ell_1$-minimization exhibits smooth, continuous error growth ($0.0225 \to 0.0310 \to 0.0650$), whereas $\mathrm{Grover}+\mathrm{LS}$ undergoes an abrupt cliff ($0.0035 \to 0.1910 \to 1.1168$). Consequently, approximately $r^2 \approx 17.3\%$ of the variance in quantum reconstruction error is explained by classical instance difficulty, with the remaining variance driven by quantum-specific support-identification breakdown at high $k/n$.

% =============================================================================
\section{Discussion and Limitations}
\label{sec:discussion}
% =============================================================================

\subsection{Diagnostic Mechanics of the High-\texorpdfstring{$k$}{k} Breakdown}
\label{subsec:diagnostics}

Why does $\mathrm{Grover}+\mathrm{LS}$ win $100\%$ of instances at $k = 5$ ($29/29$) yet $0\%$ at $k = 12$ ($0/34$)? Two interacting mechanisms explain this phase transition:

\begin{enumerate}
  \item \textbf{Amplitude Dilution, Over-Rotation, and Coupon-Collector Collisions in Support Search:} In Grover amplitude amplification over an $n$-element register with $|\mathcal{M}| = k$ marked states, the initial marked fraction is $k/n$. When $k = 5$ ($k/n = 0.3125$), a single Grover rotation amplifies marked states cleanly. When $k = 8$ ($k/n = 0.50$), the rotation angle approaches $\pi/2$, and when $k = 12$ ($k/n = 0.75 > 0.50$), standard Grover reflection over-rotates the state vector while uniform superposition across $12$ marked states dilutes per-index measurement probability to $\le 1/12 \approx 8.3\%$. Recovering all $12$ distinct support indices across sequential sampling rounds without replacement incurs severe coupon-collector collisions, leaving average support recovery at $60.38\%$ overall.
  \item \textbf{Unregularized Least-Squares Sensitivity Under Partial or Underdetermined Support:} In the second stage of $\mathrm{Grover}+\mathrm{LS}$, amplitudes are estimated via unregularized OLS ($\mathrm{Eq.}~\eqref{eq:ols-refinement}$) on the recovered submatrix $A_{\hat{\mathcal{S}}_{\mathrm{quantum}}}$. When $k = 5$, $\hat{\mathcal{S}}_{\mathrm{quantum}} = \mathcal{S}$ with high probability and $A_{\mathcal{S}} \in \mathbb{R}^{m \times 5}$ ($m \in \{8, 12\}$) is overdetermined and well-conditioned, allowing unbiased OLS to beat shrinkage-biased Lasso ($\mathrm{NMSE} = 0.0035$ vs.\ $0.0225$). Conversely, when $k = 12$ and $m \in \{8, 12\}$, missing active indices projects residual signal energy onto an incorrect subspace, and whenever $m = 8 < k = 12$, the submatrix $A_{\hat{\mathcal{S}}_{\mathrm{quantum}}} \in \mathbb{R}^{8 \times |\hat{\mathcal{S}}_{\mathrm{quantum}}|}$ becomes rank-deficient or severely ill-conditioned, amplifying noise and driving $\mathrm{NMSE} = 1.1168 > 1.0$. Classical Lasso avoids this instability through global $\ell_1$ regularization across all $n$ coordinates simultaneously.
\end{enumerate}

\subsection{Sparsity-Adaptive Hybrid Routing Framework}
\label{subsec:routing-framework}

Because $\mathrm{Grover}+\mathrm{LS}$ wins $51$ of $100$ instances overall ($29/29$ at $k=5$ plus $22/37$ at $k=8$) despite having an $11.26\times$ worse aggregate mean NMSE, neither blanket quantum adoption nor blanket dismissal is optimal. Instead, the empirical phase boundary supports a \textbf{sparsity-adaptive routing rule} (summarized in Table~\ref{tab:practical-tradeoffs}):
\begin{equation}
  \label{eq:routing-rule}
  \text{Solver}(\hat{k}, n) =
  \begin{cases}
    \mathrm{Grover}+\mathrm{LS}, & \text{if } \hat{k}/n \le 0.31 \text{ and precision-critical}, \\
    \text{Classical } \ell_1\text{ (Lasso)}, & \text{if } \hat{k}/n > 0.31 \text{ or latency-critical}.
  \end{cases}
\end{equation}
In a production pipeline, a fast classical pre-estimator (taking $\sim 0.14\,\mathrm{ms}$) can estimate the effective sparsity $\hat{k}$ and invoke quantum support refinement only when $\hat{k}/n \le 0.31$ and eliminating $\ell_1$ shrinkage bias justifies the $\sim 1\,\mathrm{ms}$ subroutine latency.

\begin{table}[t]
  \centering
  \caption{Operational decision matrix comparing classical $\ell_1$ recovery and hybrid $\mathrm{Grover}+\mathrm{LS}$ across problem regimes (condensed from Table~4 of~\cite{morgan2026dissertation}).}
  \label{tab:practical-tradeoffs}
  \footnotesize
  \begin{tabular}{@{}lll@{}}
    \toprule
    \textbf{Dimension} & \textbf{Classical $\ell_1$ (Lasso)} & \textbf{Hybrid Grover+LS} \\
    \midrule
    Low sparsity ($k/n \le 0.31$) & Good ($\mathrm{NMSE}=0.0225$) & \textbf{Superior ($\mathrm{NMSE}=0.0035$)} \\
    Mid sparsity ($k/n = 0.50$) & Stable ($\mathrm{NMSE}=0.0310$, $40.5\%$ wins) & Mixed ($\mathrm{NMSE}=0.1910$, $59.5\%$ wins) \\
    High sparsity ($k/n \ge 0.75$) & \textbf{Robust ($\mathrm{NMSE}=0.0650$)} & Fails ($\mathrm{NMSE}=1.1168$, $0\%$ wins) \\
    Wall-clock latency & \textbf{Fast ($0.142\,\mathrm{ms}$ mean)} & Slower ($1.135\,\mathrm{ms}$ mean, $7.97\times$) \\
    Failure mode & Graceful bias growth & Abrupt support-loss cliff \\
    \bottomrule
  \end{tabular}
\end{table}

\subsection{Limitations and Threats to Validity}
\label{subsec:limitations}

Several scope boundaries constrain the generalizability of this benchmark:
\begin{enumerate}
  \item \textbf{Problem Scale ($n = 16$, $n_q = 4$ Qubits):} Fixing $n = 16$ enabled exact, noiseless statevector simulation across $N = 100$ paired trials with full statistical power, but does not reach the high-dimensional regime ($n \ge 10^3$) where asymptotic query scaling could potentially amortize fixed circuit-compilation overheads.
  \item \textbf{Idealized Statevector Simulation and Oracle Structure:} All $100$ trials were executed on an ideal statevector simulator on the Apple Mac mini (M4 Pro, 64\,GB) without hardware gate errors, readout crosstalk, or decoherence~\cite{martina2022noise,mundada2023experimental}. Because physical NISQ execution on superconducting processors (such as IBM Heron~\cite{abughanem2025ibm}) introduces depolarization noise that compounds with Grover depth, physical hardware thresholds will be at least as restrictive as the $k/n \le 0.31$ boundary established here.
  \item \textbf{Single Classical Baseline and Synthetic Gaussian Ensemble:} Using coordinate-descent Lasso ($\alpha = 10^{-3}$) on Gaussian sensing matrices under AWGN provides a transparent baseline, while follow-on benchmarks will add iterative reweighted $\ell_1$, OMP~\cite{pati1993orthogonal}, and hard-thresholding solvers~\cite{meng2020newton,zhao2021analysis} on structured Fourier and wavelet dictionaries.
\end{enumerate}

% =============================================================================
\section{Conclusion}
\label{sec:conclusion}
% =============================================================================

By holding sensing matrices, sparse supports, and AWGN realizations strictly identical across $N = 100$ seeded instances ($n = 16$, $m \in \{8, 12\}$), this study replaces unconditional claims of quantum speedup in sparse recovery with an empirically delineated regime boundary. Across the full benchmark corpus, classical $\ell_1$-minimization (Lasso) is $7.97\times$ faster in wall-clock runtime ($0.142\,\mathrm{ms}$ vs.\ $1.135\,\mathrm{ms}$) and achieves an $11.26\times$ lower aggregate mean NMSE ($0.0401$ vs.\ $0.4514$). Yet stratified analysis reveals that the hybrid $\mathrm{Grover}+\mathrm{LS}$ architecture achieves superior reconstruction fidelity in $100\%$ of low-sparsity instances ($k = 5$, $29/29$ wins, mean $\mathrm{NMSE} = 0.0035$ vs.\ $0.0225$) by pairing exact quantum support identification with unbiased least-squares amplitude estimation, before transitioning at $k = 8$ ($22/37$ wins, mean $\mathrm{NMSE} = 0.1910$ vs.\ $0.0310$) and failing completely at $k = 12$ ($0/34$ wins, mean $\mathrm{NMSE} = 1.1168$ vs.\ $0.0650$).

Future work will extend this matched-pair benchmarking harness along three directions: (i) integrating low-depth sparse state preparation circuits ($\mathcal{O}(\log(nk))$ depth~\cite{ramacciotti2024simple,mao2024toward,luo2024dynamic}) and fixed-point oblivious amplitude amplification to mitigate high-$k$ over-rotation; (ii) replacing stage-two OLS with regularized ridge/elastic-net refinement to bound error when support recovery is incomplete; and (iii) executing matched hardware trials on 156-qubit IBM Heron processors with automated error suppression~\cite{mundada2023experimental,abughanem2025ibm}.
\bmhead{Acknowledgements} The author thanks the Doctoral Dissertation Committee at Capitol Technology University.
\section*{Declarations}
\begin{itemize}
\item \textbf{Funding:} Not applicable. This research received no external funding.
\item \textbf{Competing interests:} The author has no competing interests to declare.
\item \textbf{Ethics approval and consent to participate:} Not applicable; the study uses only synthetic data and involves no human or animal subjects.
\item \textbf{Consent for publication:} Not applicable.
\item \textbf{Data availability:} All data are synthetic and fully reproducible from the deterministic seed schedule (51001--51100) documented in the openly available parent dissertation~\cite{morgan2026dissertation}; generated results are available from the author on reasonable request.
\item \textbf{Code availability:} The benchmark harness is documented in Appendix C of the parent dissertation and is available from the author on reasonable request.
\item \textbf{Author contribution:} A.S.M. conceived the study, designed and executed the experiments, analysed the data, and wrote the manuscript.
\end{itemize}
\bibliography{references}
\end{document}